\documentclass[structabstract]{aa}  
\usepackage{graphicx}
\usepackage{txfonts}
\begin{document}
\title{Bolometric luminosity variations in the luminous blue variable \object{AFGL2298}
\thanks{Based on observations made at the European Southern Observatory, Paranal, Chile under programs
077.C-0207 and 079.D-0769.}}
\author{J. S. Clark\inst{1}
\and P. A. Crowther\inst{2}
\and V.~M.~Larionov\inst{3,4}
\and  I. A. Steele\inst{5}
\and B. W. Ritchie\inst{1,6}
\and A.~A.~Arkharov\inst{7}}
\institute{
$^1$Department of Physics and Astronomy, The Open 
University, Walton Hall, Milton Keynes, MK7 6AA, UK\\
$^2$ Department of Physics \& Astronomy, University of Sheffield, Sheffield, S3 7RH, UK\\ 
$^3$ Astronomical Institute of St. Petersburg University, Petrodvorets, Universitetsky pr. 28, 198504 
St. Petersburg, Russia\\
$^4$ Isaac Newton Institute of Chile, St. Petersburg Branch\\
$^5$ Astrophysics Research Institute, Liverpool JMU, Twelve Quays House, Egerton Wharf, Birkenhead, CH41, 1LD, UK\\
$^6$ IBM United Kingdom Laboratories, Hursley Park, Winchester, Hampshire, S021 2JN, United Kingdom.\\
$^7$ Pulkovo Astronomical Observatory, 196140 St. Petersburg, Russia
}

\abstract{}{We characterise the variability in the physical properties of the 
luminous blue variable \object{AFGL 2298} (IRAS 18576+0341) between 1989 -- 2008.}
{In conjunction with published data from 1989-2001, we have undertaken a 
long term (2001 -- 2008) near-IR spectroscopic and photometric 
observational 
campaign for this star and utilise a non-LTE model atmosphere code to 
interpret these data.}
{We find \object{AFGL 2298} to have been highly variable during the two 
decades covered by the observational datasets. Photometric variations of 
$\geq$1.6 mag have been observed in the JHK wavebands; however, these are 
not accompanied by correlated changes in near-IR colour. Non-LTE model 
atmosphere analysis of 4 epochs of K band spectroscopy obtained between 
2001-7 suggests that the photometric changes of \object{AFGL 2298} were 
driven by expansion and contraction of the stellar photosphere accompanied 
by comparatively small changes in the stellar temperature 
($\Delta$T$_{\ast}\sim$4.5~kK). Unclumped mass loss rates throughout this 
period were modest and directly comparable to those of other highly 
luminous (candidate) LBVs. However, the main finding of this analysis was 
that the bolometric luminosity of \object{AFGL 2298} appears to have 
varied by at least a factor of $\sim$2 between 1989-2008, with it  being one 
of the most luminous stars in the Galaxy during maximum. Comparison to 
other LBVs that  have undergone non bolometric luminosity conserving `eruptions' 
shows such events to be heterogeneous, with \object{AFGL 2298} the 
least extreme example. These results - and the diverse nature of both the 
quiescent LBVs and associated ejecta - may offer support to the suggestion 
that more than one physical mechanism is responsible for such behaviour.}
{}


\maketitle

\section{Introduction}

Luminous blue variables (LBVs) represent a transitional state in the 
evolution of massive stars between main sequence and hydrogen depleted 
Wolf-Rayets and are characterised by significant photometric and 
spectroscopic variability (e.g. Lamers \cite{lamers}, Humphreys \& 
Davidson \cite{HD}). Two characteristic modes have historically been 
identified: (i) 1-2 magnitude excursions on $\sim$year timescales at 
constant bolometric luminosity (L$_{\rm bol}$) and (ii) giant ($\geq$2mag)
eruptions, during which the luminosity of the star increases, but for 
which the timescales are currently uncertain due to their rarity.

Recently, LBVs have been the subject of renewed interest. With the likely 
downwards revision of Main Sequence mass loss rates it has been 
suggested that \object{$\eta$ Car}-like giant eruptions play a key role 
in stripping the H rich mantle from post-MS stars prior to the WR phase 
(Smith \& Owocki \cite{smith06}). Moreover, some anomalously faint Type 
IIn SNe have been hypothesised to be LBVs in an eruptive phase rather than 
core-collapse events (e.g. Humphreys et al. \cite{h99}, Goodrich et al. 
\cite{goodrich}). Finally, several lines of evidence have been advanced to 
suggest that LBVs may be the immediate progenitors of a subset of Type II 
SN, including some of the most luminous Type IIn events ever observed 
(e.g. Smith et al. \cite{smith07}, Gal-Yam et al. \cite{galyam}, Trundle 
et al. \cite{trundle}, Pastorello et al. \cite{pastorello}, Kotak \& Vink 
\cite{kotak}), in turn 
leading to the suggestion that the occurrence of \object{$\eta$ Car}-like 
eruptive events may warn of an imminent SN (e.g. Smith et al. 
\cite{LBVSN}, Woosley et al. \cite{woosley}).

Given these possibilities, an understanding of the physics leading to both 
modes of LBV variability is urgently required. Unfortunately, such a goal 
is hampered by the evident rarity of LBVs and their attendant outbursts. 
Only two examples of non L$_{\rm bol}$ conserving eruptions have been 
spectroscopically studied -  HD~5980 (Koenigsberger 
\cite{koenigsberger}) 
and NGC2363-V1 (Drissen et al. \cite{drissen}) - both of which occurred in 
low metallicity environments. However, with the expansion of IR astronomy 
over the past decade, a number of new candidate LBVs have been 
identified within the Galaxy via both imaging and spectroscopy (e.g. Clark 
et al. \cite{clark05} and references therein). Of these, the presence of 
high mass circumstellar nebulae surrounding a subset of B hypergiants is 
particularly suggestive of enhanced mass loss in the past, possibly 
associated with giant eruptions (Ueta et al. \cite{ueta}, Clark et al. 
\cite{clark03b}, Smith \& Owocki \cite{smith06}).

\begin{figure*}
\resizebox{\hsize}{!}{\includegraphics[angle=0]{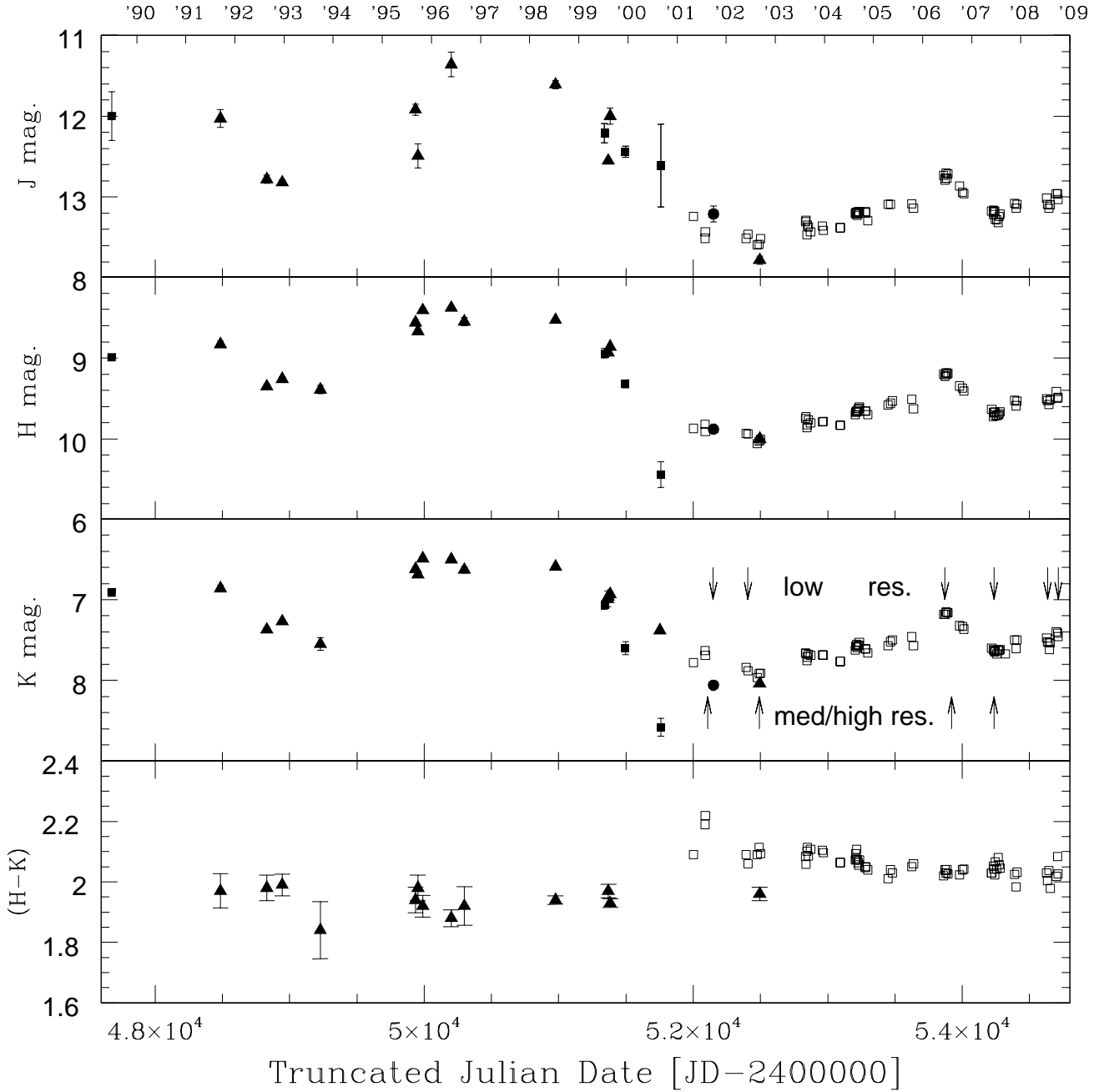}}
\caption{JHK band lightcurves and (H-K) colour index for \object{AFGL 2298} spanning $\sim$two decades. Data are from 
Jimenez-Esteban et al. (\cite{jimenez}; filled triangles), 
Ueta et al. (\cite{ueta}; filled squares), Pasquali \& Comeron (\cite{pasquali}; 
open circles) and Clark et al. (\cite{clark05}) and this work (open squares). 
Where not shown, error-bars are smaller than the size of the respective symbols.
The arrows in the K band  panel indicate the times at which the spectroscopic observations were made, 
with the low, medium and high  resolution observations made with the 
AZT-24 1.1m, UKIRT/CGS4 and the VLT/ISAAC respectively (see Sect. 2).} 
\label{light_curve}
\end{figure*}

\object{AFGL 2298} (=\object{IRAS  18576+0341}), observed as part of a 
long term spectroscopic and photometric monitoring  campaign of 
(candidate) LBVs, has been found to be highly variable (Clark et 
al. \cite{clark03a}).  Spectroscopic analysis of near-IR observations
suggested significant changes in stellar radius and temperature 
between 2001-2002, indicative of an LBV excursion. In this paper we 
update the results of this effort, encompassing an extended lightcurve 
and additional medium resolution spectroscopy and an accompanying 
analysis with the non-LTE model atmosphere code CMFGEN (Hillier \& Miller 
\cite{hil98},  \cite{hil99}).  Finally, in light of these results and 
other recent  results, we discuss the observational properties of LBV 
variability and  the implications for the formation of ejection nebulae.

\section{Data Reduction}

Near-IR JHK broadband photometric observations of \object{AFGL 2298} were 
obtained at the AZT-24 1.1m telescope in Campo Imperatore (Italy) from 
2001 March - 2008 October.  The SWIRCAM 256x256 HgCdTe detector was 
employed, yielding a scale of 1.04~arcsec/pix, resulting in a 
$\sim4^{\prime} \times 4^{\prime}$ field of view. Standard techniques of 
sky subtraction and flat-fielding were applied. Between two and four 
standards were employed for the calibration of the photometry. All 
standards were located within the target frames, had near-IR magnitudes 
between 9.9 and 11.5~mag. and were found to be constant to within 
$\pm$0.02mag over the course of the observations. The resultant lightcurve 
is presented in Fig.~\ref{light_curve}.

Spectra of \object{AFGL 2298} have been obtained from a number of 
different telescopes and are summarised in Table~\ref{log}, and presented 
in Figs.~2--\ref{hires}. A description of the UKIRT/CGS4 observations and reductions may be 
found  in Clark et al. (\cite{clark03a}). Long term, low resolution 
($R\sim$270) 
spectroscopic monitoring was undertaken with the AZT-24 1.1m telescope and 
the IR imaging camera SWIRCAM+HK band grism - providing spectral coverage 
between 1.45 -- 2.38$\mu m$. Higher resolution (R$\sim$8900) follow up 
observations were subsequently made with the VLT in 2006 and 2007.

The VLT observations were made with ISAAC in the  short-wavelength 
(SW) medium resolution (MR) mode with a narrow 0.3'' slit. To 
achieve 
spectral coverage from 2.04 -- 2.22$\mu$m two exposures were 
obtained, centred at 2.10$\mu$m and 2.16$\mu$m. All data were taken with a 
count rate of below 10,000 ADU, therefore no correction for non-linearity 
was necessary. Data reduction was accomplished using the {\sc FIGARO} and 
{\sc KAPPA} software packages. Note that the telluric correction was poor 
for the 2007 spectrum due to rapidly varying sky conditions.

\section{Results}

\subsection{Photometry}

With the addition of the data presented here and in Jimenez Esteban et al. 
(\cite{jimenez}) the lightcurve for \object{AFGL 2298} extends for $\sim$2 
decades, revealing significant near-IR variability throughout this time 
(Fig. 1). Our well sampled lightcurve from 2001-8 reveals two minima. The 
first, deeper minimum occurred in the second half of 2002 and the second in 
the first half of 2007, with the corresponding maximum in mid 2006. Low 
amplitude ($\sim$0.2mag) variability over timescales of $\sim$days is 
superimposed on the long term trends in all 3 wavebands.

In addition to these minima, the comparatively poorly sampled data set of 
Jimenez Esteban et al. (\cite{jimenez}) provides evidence for a further 
photometric minimum between 1993-5 and a maximum between 1996--1999. The 3 
minima and 2 maxima appear to have differing intensities, with a maximum 
peak to trough variation of $\geq$1.6 mag observed between the 1996--1999 
maximum and 2002 minimum.

\begin{table}
\begin{center}
\caption[]{Log of the spectroscopic observations 
between 2001-8.}\label{log}
\begin{tabular}{cccr}
\hline
\hline
Date &  Telescope & $\lambda$ & Resolution \\
     &                 & ($\mu$m) & \\
\hline
27/06/01  & UKIRT/CGS4 & 2.0-2.5 & 800\\
22/07/02  & AZT-24 1.1m  & 1.45-2.38 & 270 \\
06/08/02  & UKIRT/CGS4 & 2.0-2.5 & 800\\
17/05/06  &  AZT-24 1.1m  & 1.45-2.38 & 270 \\
30/06/06  & VLT/ISAAC & 2.04-2.22 & 8900 \\ 
15/05/07 &  AZT-24 1.1m  & 1.45-2.38 & 270 \\
09/05/07 & VLT/ISAAC & 2.04-2.22 & 8900 \\ 
16/06/08 & AZT-24 1.1m  & 1.45-2.38 & 270 \\
04/09/08 & AZT-24 1.1m  & 1.45-2.38 & 270 \\
\hline
\end{tabular}
\end{center}
\end{table}

\begin{figure}
\includegraphics[width=8cm]{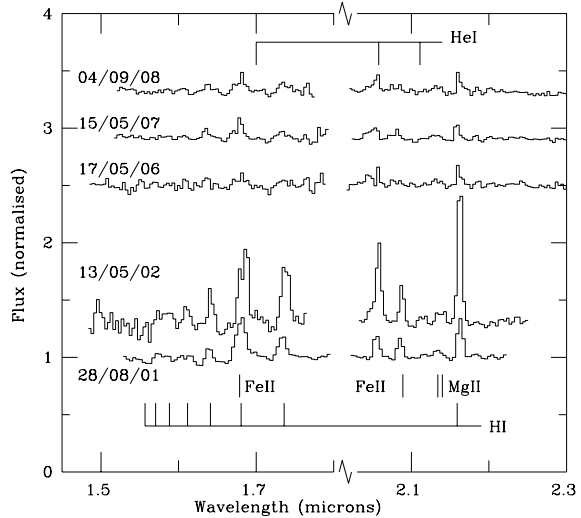}\\
\caption{Selected low resolution AZT-24  spectra of \object{AFGL 2298} from 2002-2008. Note 
the significant reduction  in the strengths of all emission lines between 2006-9 in comparison
to 2001-2 (also see Fig. 3).}
\end{figure}

 \begin{figure*}
\includegraphics[width=12cm]{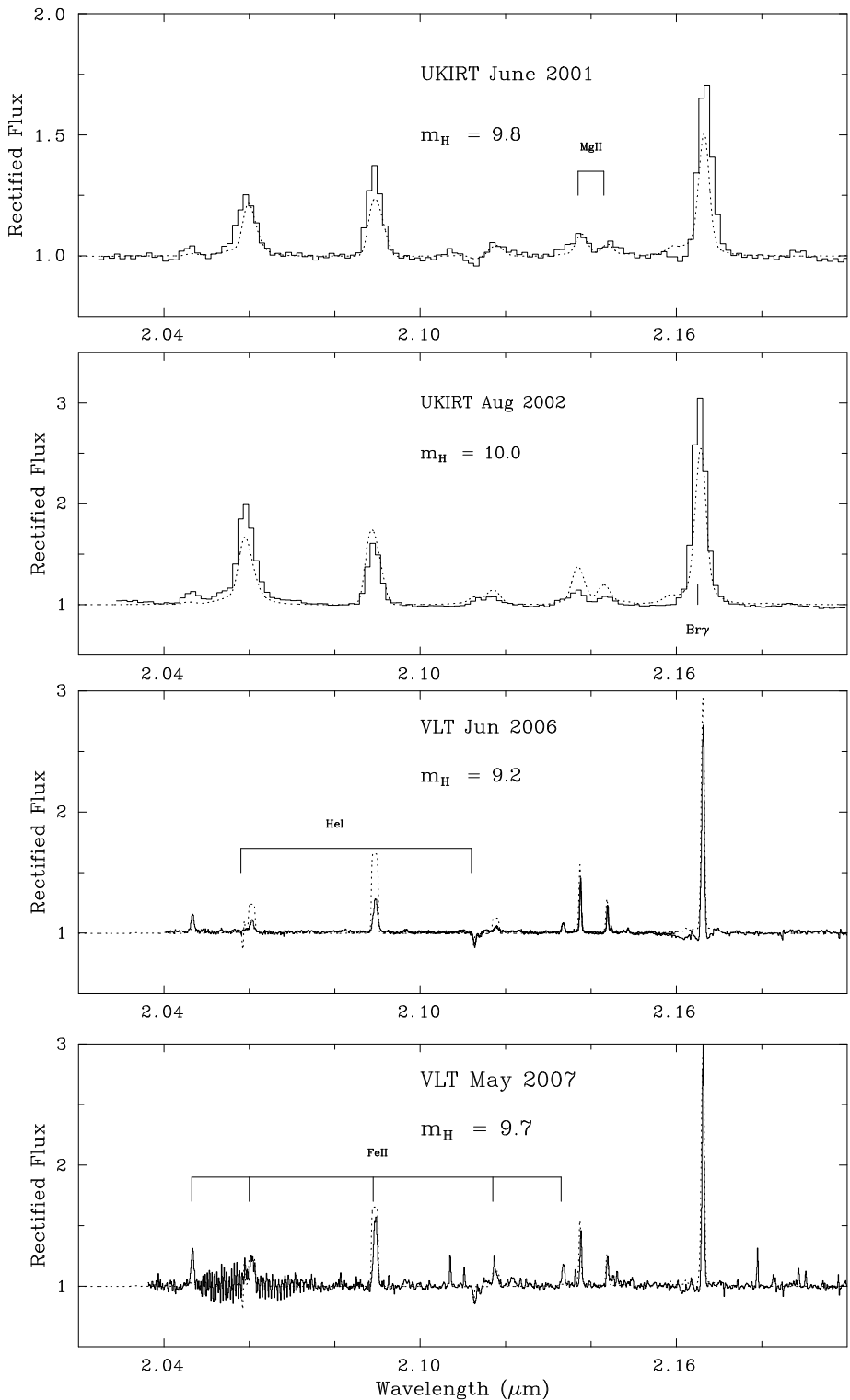}\\
\caption{Montage of 4 epochs of medium (UKIRT/CGS4: R$\sim$800) and high 
(VLT/ISAAC; R$\sim$8900) resolution  K band 
spectroscopy  of \object{AFGL 2298}  from 2001 and 2007 (solid lines) with the major transitions indicated (note that telluric subtraction was poor for the 2007 observation). The resultant spectra from the non-LTE model atmosphere 
analysis in Sect. 4.2 are overplotted (dotted lines). Note the different scales on the y-axis.}\label{hires}
\end{figure*}

In the bottom panel of Fig.~\ref{light_curve} we plot the (H-K) colour 
index, finding no 
systematic variation with either the long term secular changes or short 
timescale low amplitude variability during these observations. As such 
this behaviour differs from that demonstrated by the candidate LBVs 
\object{G24.73+0.69} and \object{G26.47+0.02} (Clark et al. 
\cite{clark05}, Clark et al. in prep.), where both stars became 
significantly redder as they brightened.  Indeed the lack of colour 
variability associated with either rising or falling branches of the 
lightcurve of \object{AFGL 2298} argues against the changes being due to 
episodes of enhanced mass loss\footnote{Where a brightening accompanied by 
an increase in reddening may reflect increased emission from circumstellar 
material (and vice versa e.g. \object{WR137} and \object{WR19}; Williams 
et al. \cite{williams} \& Veen et al. \cite{veen}), while a reduction 
(increase) in brightness accompanied by an increase (decrease) in 
reddening may be attributed to increased (decreased) extinction due to 
very dense circumstellar material (e.g. \object{IRAS 16029-3041}; 
Jimenez-Esteban et al. \cite{jimenez}).}. Instead, we attribute it to 
changes in the properties of the underlying star, although the 
identical near-IR colours of early OB stars preclude a determination 
of stellar temperature from these data alone.

 For our spectroscopic analysis, we adopt $E(B-V)$=9 mag from Ueta et 
al. 
(\cite{ueta}), equating to an infrared extinction of $A_{\rm J}=8.25$, 
$A_{\rm H}=4.65$ and $A_{\rm K}=2.7$, plus a distance of 10 kpc (distance 
modulus 15.0 mag), as in Clark et al. (\cite{clark03a}). With this choice
of infrared extinction, absolute magnitudes in the J,  H and K bands
differ by at most 0.1 mag, in agreement with predictions from atmospheric
models (Section 4). No  evidence was found for hot dust in any of 
the 4 epochs of spectroscopic observations -- JHK magnitudes were 
well reproduced by the reddened spectral energy distributions in all 
cases.

\begin{figure} \includegraphics[angle=-90,width=7cm]{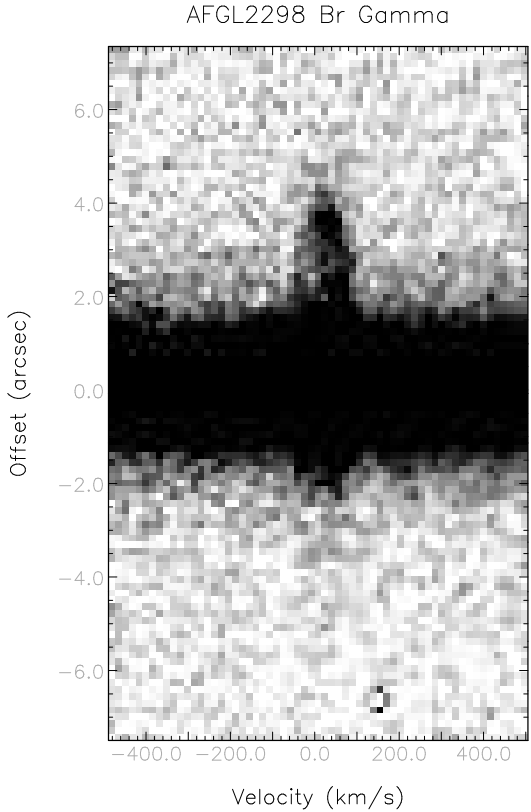}\\ 
\caption{Long slit spectrum of \object{AFGL2298} centred on the Br$\gamma$ 
line with a North(uppermost)-South orientation. Extended, nebular emission 
is present up to 4~arcsec northwards of  the star (0.2~pc for a distance of 
10~kpc), with the elliptical morphology consistent with material expanding 
at up to $\pm\sim$70~kms$^{-1}$. Note that the physical extent and the 
pronounced asymmetrical nature of the emission is consistent with the 
morphology of the ionised component of the nebula found from the radio 
continuum observations of Umana et al. (\cite{umana}). }\label{nebula} 
\end{figure}

\subsection{Spectroscopy}

Figure~2 presents our AZT-24 low resolution (R$\sim$270) H-
and  K-band spectroscopy of AFGL 22298, while low-medium resolution
(R$\sim$800-8900) UKIRT and VLT K-band spectroscopy is 
presented in Figure~\ref{hires}. These illustrate the significant 
change in spectral morphology between 2001 - 2002 reported by Clark et al. 
(\cite{clark03a}), as well as a subsequent dramatic reduction in strength 
of the emission lines of all species by 2006 May, a state which has 
persisted for the past $\sim$2.5~yr. Unfortunately, a lack of observations 
between 2002 August and 2006 May prevent us localising the transition - if 
indeed it were a sudden event - over the course of the secular brightening 
of \object{AFGL 2298} during this period (recall Fig~\ref{light_curve}). 
However, 
following the 2006 maximum, no pronounced changes are observed in the low 
resolution spectra despite the subsequent photometric variability.

This is emphasised by the  medium resolution VLT/ISAAC spectrum of 
2006 
June 30 (Fig.~\ref{hires}) - fortuitously obtained close to infrared 
maximum (Fig.~\ref{light_curve}) - which 
reveals a  narrow emission line spectrum (FWHM$_{Br \gamma}$=70kms$^{-1}$) 
dominated 
by Br$\gamma$ and low excitation metallic species. Comparison to the 2002 
UKIRT/CGS4 spectrum obtained during the photometric minimum confirms the 
pronounced weakening in line emission for all species; He\,{\sc i} emission 
- prominent in the 2001--2002 spectra - is entirely absent, with the 
He\,{\sc i} 2.112$\mu$m transition seen in absorption.

A second  VLT/ISAAC medium resolution spectrum obtained on 2007 May 9 
- 
this time  during the subsequent local photometric minimum - shows a 
moderate 
recovery in the strength of Fe\,{\sc ii} emission lines, but no significant 
changes in the strength of Br$\gamma$ or the Mg\,{\sc ii} and He\,{\sc i} 
lines. These changes - in particular the absence of He\,{\sc i} emission - 
suggest a modest reduction in temperature after 2002 (Sect. 4.2; 
Table~\ref{summary_afgl}), 
although {\em all the spectra are entirely consistent with the 
classification of AFGL 2298 as a cool highly luminous B supergiant 
throughout the 2001-9 period.}

Finally, these observations reveal spatial extended emission in the 
Br$\gamma$ line at distances of up to 4~arcsec from the star, allowing us 
to infer an expansion velocity of $\pm \sim$70kms$^{-1}$ for the detached 
ejection nebula associated with \object{AFGL 2298} (Fig.~\ref{nebula}). 
Adopting the 
radial extent and mass given by Ueta et al. (\cite{ueta})  implies a 
kinematic age of 8300~yr and consequently a {\em time averaged} mass loss 
rate of 1.2$\times10^{-3}$ M$_{\odot}$yr$^{-1}$ during nebular formation.

\section{Physical properties of \object{AFGL 2298}:}

\subsection{1989--2001}
 
In the absence of spectroscopic data, the interpretation of the 
photometric lightcurve of \object{AFGL 2298} prior to 2001 is subject to 
uncertainties due to the fact that near-IR colours of hot OB stars are 
$\sim$constant. Nevertheless, given the similar stellar luminosities, 
magnitude of the excursions ($\Delta$J$\sim$2.5; Fig.~\ref{light_curve} \& 
Groh, priv. 
comm.)  and timescale of variability ($\sim$2-4yrs; 
Fig.~\ref{light_curve}, Spoon et al. 
\cite{spoon}), it would appear that \object{AG Car} should serve as a 
suitable template for comparison to \object{AFGL 2298}. Groh 
(\cite{thesis}) studied \object{AG Car} through two full photometric 
cycles from 1986-2005 and found a transition from T$_{\rm eff} \sim$24 to 
$\sim$8 kK, anti-correlated with both V and J band magnitudes, as expected 
for LBV excursions.

\begin{table}
\begin{center}
\caption{Summary of physical and wind properties of AFGL 2298 for
2001--2007.}\label{summary_afgl}
\begin{tabular}{lccccc}
\hline
\hline
Epoch &  2001 Jun &  2002 Aug &    2006 Jun &     2007 May \\
\hline
J     &    13.4   &   13.7    &      12.7   &      13.2    \\
H     &    9.8    &   10.0    &       9.2   &       9.6    \\ 
K     &    7.9    &    7.9    &       7.2   &       7.6    \\ 
%M$_{\rm J}$   & -9.8  & -9.55  &    -10.55  &     -10.05 \\
M$_{\rm H}$   & -9.85 & -9.65  &    -10.45  &     -10.05 \\
%M$_{\rm K}$   & -9.8  & -9.8   &    -10.5   &     -10.1  \\
\hline
T$_{\ast}$ (kK)& 12.5  & 15.5   &     11.0   &      11.5  \\
R$_{\ast}$ (R$_{\odot}$)& 262   &     158  &    385 &   312 \\
T$_{\rm eff}$  (kK)  &  11.7 &  10.9 &   10.3 &    10.9  \\
R$_{2/3}$  (R$_{\odot}$)&  300  &   320 &    444 &  353 \\
BC$_{\rm H}$  &  -0.85  &       -0.90  &          -0.55  &        -0.65 \\
L$_{\ast}$/L$_{\odot}$ &   1.5$\times$10$^6$ & 1.3$\times$10$^6$ & 2.0$\times$10$^6$ & 1.5$\times$10$^6$ \\
\.{M} (M$_{\odot}$yr$^{-1}$) & 4.5$\times 10^{-5}$ & 1.2$\times 10^{-4}$ & 5.2$\times 10^{-5}$  & 4$\times 10^{-5}$ \\
v$_{\infty}$ kms$^{-1}$  & {\em 200} & {\em 200} & 125  & 125 \\
H/He   &   {\em 5} & {\em 5} & {\em 5} & {\em 5} \\
\hline
\end{tabular}
\note{For all epochs we have assumed DM=15,
E(B-V)=9, $A_{\rm J}=8.25$, $A_{\rm H}=4.64$, $A_{\rm K}=2.69$ (from 
Ueta et al. \cite{ueta})
and H/He=5.0, while 
v$_{\infty} \sim$200kms$^{-1}$ was assumed for 2001--2002; these 
latter two parameters are given in italics. Mass-loss rates presented here are 
{\it not}  corrected for wind clumping. We find errors of $\pm$2 kK 
($\pm$0.2 mag in BC),  $\pm$1.5 kK ($\pm$0.2mag) and  $\pm$0.5 kK 
($\pm$0.1mag) for 2001, 2002 and 2006-2007 respectively.}
\end{center}
\end{table}

Somewhat unexpectedly, in light of the prevailing orthodoxy,  Groh et 
al. 
(\cite{groh}) found that the L$_{\rm bol}$ of AG Car was {\em not} 
conserved over the period of these cycles, with a reduction of $\sim$50\% 
during photometric maximum/temperature minimum when compared to 
photometric minimum/temperature maximum. Similar behaviour has been 
observed for \object{S Dor} (e.g. van Genderen \cite{vG}), with Lamers 
(\cite{lamers95}) interpreting the resultant deficit in radiative energy 
during photometric maximum as being due to the increased mechanical energy 
required to move and subsequently support the outer layers of the star 
against gravity at a greatly expanded radius.

However, \object{AFGL 2298} was significantly cooler (T$_{eff} \sim$10.8 kK) in the 
2002--2003 photometric minimum than \object{AG Car} in a comparable phase 
($\sim$24 kK). Such a low temperature implies a (H-band) bolometric 
correction of $<$1 mag (Sect. 4.2). If \object{AFGL 2298} were to have 
mirrored the behaviour of AG Car between 1989-2001 - i.e. with periods of 
increasing IR brightness corresponding to a decrease in temperature (and 
vice versa) - the magnitude of near IR variability observed ($>$1.6 mag) 
would preclude such changes occuring at reduced or even constant 
L$_{\rm bol}$ 
and instead would indicate that the increase in IR brightness that 
accompanied a reduction in stellar temperature {\em would at least in part 
be due to a genuine increase in luminosity.} Trivially, such a conclusion 
would also be drawn if the variability occurred at constant or increasing 
temperature. We will return to this below, upon consideration of the 
results of non-LTE modeling of the post 2001 data.

 \begin{figure}
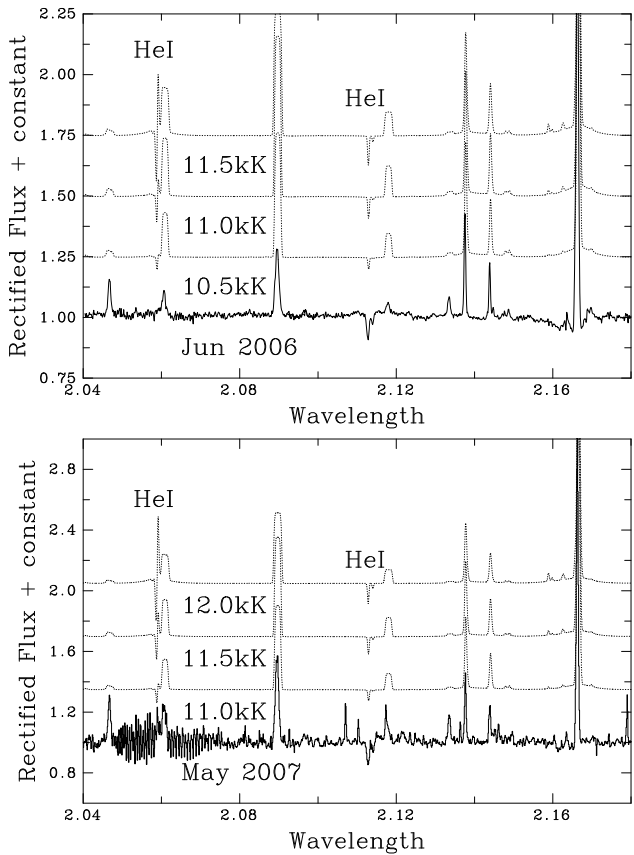
 
\includegraphics[width=0.625\columnwidth,angle=270]{12358fg5.eps}\\ 
\includegraphics[width=0.625\columnwidth,angle=270]{12358fg6.eps}\\ 
\caption{VLT/ISAAC spectroscopy of AFGL  2298 for (upper panel) Jun 
2006
together with synthetic spectra for 11$\pm$0.5 kK; (lower panel) May 2007
together with synthetic spectra of 11.5$\pm$0.5 kK, demonstrating the 
temperature sensitivity of K-band He\,{\sc i} lines.}\label{fits}
\end{figure}

\subsection{2001--2007}

While we may not unambiguously interpret the 1989--2001 behaviour of 
\object{AFGL 2298} from photometric data alone, the contemporaneous 
spectroscopy in the period 
2001--2007 allows us to accomplish this via a non-LTE  spectroscopic 
analysis using 
CMFGEN (Hillier \& Miller \cite{hil98}, \cite{hil99}).
We {\em explicitly} follow an identical 
methodology to that employed in Clark et al. (\cite{clark03a}), in order 
to permit a direct comparison of all four epochs of spectroscopic 
observations.

\subsubsection{Atmospheric code}

 CMFGEN solves the 
radiative transfer equation in the co-moving 
frame, under the additional constraints of statistical and radiative 
equilibrium. It does not solve the momentum equation, so a density or 
velocity structure is required. For the  supersonic part, the velocity is 
parameterized with a classic $\beta$-type  law. CMFGEN incorporates line 
blanketing through a super-level  approximation, in which atomic levels of 
similar energies are grouped into 
a single super-level which is used to compute the atmospheric structure. 
Our atomic model is similar to that employed by Drissen et al. 
(\cite{drissen}), including ions from H\,{\sc i}, He\,{\sc i-ii}, N\,{\sc 
i-ii}, O\,{\sc i-ii}, Mg\,{\sc i-ii}, Al\,{\sc ii-iii}, Si\,{\sc ii-iii}, 
S\,{\sc ii-iii}, Ti\,{\sc ii-iii} and Fe\,{\sc ii-iii}. By number, the 
main contributors to line blanketing are Fe\,{\sc ii-iii}. H/He$\sim$5 by 
number was adopted for all models, since solely K-band datasets do not 
readily allow the determination of H/He contents without a temperature 
diagnostic independent of helium. Solar abundances are taken for metals, 
with the exception of 5 times solar for nitrogen, as expected for 
chemically evolved LBVs.

Stellar temperatures, $T_{\ast}$, correspond to a 
Rosseland  optical depth of 20, which are typically (up to) a few hundred
degrees  higher than effective temperatures, $T_{2/3}$, relating to 
a Rosseland optical depth of 2/3.

 We have assumed a depth-independent Doppler profile for all lines 
when 
solving for the atmospheric structure in the co-moving frame, while in the 
final calculation of the emergent spectrum in the observer's frame, we 
have adopted a uniform turbulence of 20 km\,s$^{-1}$. Incoherent electron 
scattering and Stark broadening for hydrogen and helium lines are 
adopted.

One improvement afforded by the ISAAC datasets was the means of 
observationally constraining the terminal wind velocity in 2006-7, which 
had to be estimated for the 2001--2002 data.  As in Clark et al. 
(\cite{clark03a}), a $\beta$=2 velocity law was adopted for all 
subsequent epochs.  The form of the velocity law is not expected to differ 
from epoch to epoch, especially between the crucial 2006 and 2007 
datasets. With regard to wind clumping, this was {\em not} included for 
consistency  with Clark et al. (\cite{clark03a}), although volume filling 
factors of $f\sim$0.1 typically reduce mass-loss rates by a factor of 
$\sqrt{1/f}\sim$3.

\subsubsection{Analysis}

We derive the stellar temperature of AFGL 2298 primarily using the 
K-band He\,{\sc i} line diagnostics. Specifically, the higher resolution 
2006-7 VLT/ISAAC datasets 
allowed the He\,{\sc i} 2.058$\mu$m emission line and 2.112$\mu$m 
absorption line to be resolved, thereby allowing significantly more 
robust physical parameters (e.g. He\,{\sc i} 2.058$\mu$m and Fe\,{\sc ii} 
2.060$\mu$m were blended in 2001--2002 UKIRT/CGS4 datasets).

For Jun 2006, the simultaneous presence of strong 2.112$\mu$m 
absorption 
plus negligible 2.058$\mu$m emission occurs for a narrow range in 
temperatures,  namely 11$\pm$0.5 kK (equivalent to $\pm$0.1 mag in 
Bolometric Correction;  BC), as shown in the upper panel of 
Fig.~\ref{fits}. Similar conclusions
were drawn for two LBVs in the Quintuplet cluster by Najarro et al. 
(\cite{paco08}).
The mass-loss rate is  fixed from Br$\gamma$ with other 
K-band  features either reproduced  satisfactorily (Mg\,{\sc ii} doublet), 
or predicted somewhat too strong (Fe\,{\sc ii} lines).

For May 2007, weak 2.112$\mu$m absorption plus weak 2.058$\mu$m P 
Cygni 
emission implies a slightly higher stellar temperature of 11.5$\pm$0.5 kK 
(again reliable to $\pm$0.1 mag in BC) -- see lower panel of 
Fig.~\ref{fits} -- plus a reduced mass-loss rate from  Br$\gamma$. For 
this epoch, both Mg\,{\sc ii}  and Fe\,{\sc ii} lines are now 
satisfactorily  reproduced, including the  Fe\,{\sc ii} feature partially 
blended with 2.058$\mu$m. 

For reference, we have also investigated the potential effect of 
a varying 
$\beta$ velocity law, for the 2007 epoch. Use of a $\beta$ = 1 
law would favour a 0.5~kK higher temperature, yet maintain the H-band 
bolometric correction to within 0.03 mag. Similarly, a $\beta$ = 3 law
favours a 0.5~kK lower temperature, albeit with a H-band bolometric
correction consistent to within 0.05 mag. These differences lie within the 
formal uncertainties quoted above, and so do not affect our
main conclusions. Alas, no diagnostics
are available in the K--band from which a determination of the form of the 
velocity law can be uniquely obtained.

For the 2001 and 2002 epochs, we have retained models from Clark et 
al. (\cite{clark03a}), but in view 
of the unresolved He\,{\sc i} + Fe\,{\sc ii} 2.06$\mu$m blend and weak 
He\,{\sc i} 2.112$\mu$m feature we admit substantially larger 
uncertainties in temperatures - 12.5$\pm$1.5 kK for Jun 2001 and 15$\pm$2 kK 
for Aug 2002 ($\pm$0.2 mag in BC). In view of the limitations outlined 
below, it is  apparent that  the comparison 
between physical parameters for AFGL 2298  for 2006 and 2007  epochs is 
more significant than that between 2001 and 2002 or between UKIRT/CGS4
and VLT/ISAAC datasets.

 Specifically for Jun 2001, the blended emission feature at 2.06$\mu$m 
is significantly stronger than the Fe\,{\sc ii} 2.04$\mu$m line, 
and indeed comparable  in strength to Fe\,{\sc ii} 2.089$\mu$m, suggesting 
that P Cygni
He\,{\sc i} 2.058$\mu$m emission makes a significant contribution to the
feature. In contrast He\,{\sc i} 2.112$\mu$m remains weakly in absorption
(albeit blended with Fe\,{\sc ii} 2.12$\mu$m), from which we estimate
a temperature of 12.5$\pm$1.5 kK (recall Fig.~\ref{fits}). Errors are 
calculated on the basis that 2.06$\mu$m is dominated by Fe\,{\sc ii} at 
significantly lower temperatures,  while negligible 2.112$\mu$m He\,{\sc 
i} absorption is predicted at higher temperatures. 

 For Aug 2002, a higher stellar 
temperature of 15.5$\pm$2  kK is suggested on the basis that the 
2.06$\mu$m blend is now stronger in emission than 2.089$\mu$m Fe\,{\sc 
ii}, such that it is presumably dominated by He\,{\sc  i} 2.058$\mu$m. In 
addition, He\,{\sc i} 
2.112$\mu$m is now observed to be weakly in emission.  Similar issues 
apply regarding errors  for the 2002 dataset, although a slightly
higher temperature (by $\sim$0.5~kK) is potentially  favoured by the 
weakness of  the Mg\,{\sc ii} 2.13--2.14$\mu$m doublet.

\subsection{ Results}

 A summary of spectroscopic results is presented in 
Table~\ref{summary_afgl}. We find that the large changes in photometric 
magnitude 
between 2001--2007 (e.g. $\Delta$J=1.0 mag) were driven by significant 
variation in  stellar radius (from 160-385 R$_{\odot}$ between 
2002--2006), albeit with only 
moderate changes in temperature  ($\Delta$T$_{\ast}\leq$4.5 kK);   
significantly smaller than observed in the S Dor excursions of \object{AG 
Car} (Groh et al. \cite{groh}). During this period, the mean 
photospheric expansion velocity was $v_{exp}$=1.25~kms$^{-1}$.

In particular, the photometric fading between 2006--2007 appears to result 
almost entirely  from a reduction in stellar radius from 385 to 310 
R$_{\odot}$ 
($v_{exp}$=-1.6~kms$^{-1}$) at near-constant temperature  
($\Delta$T$_{\ast}\sim0.5$ kK). {\em This combination of parameters 
results 
in a decrease in L$_{\rm bol}$ for \object{AFGL 2298} from 2 to 
1.5$\times10^6$L$_{\odot}$ in less than one year.} Such a 
variation differs from that inferred for both \object{S Dor} (Lamers 
\cite{lamers95}) and \object{AG Car} (Groh et al. \cite{groh}) in 
the 
sense  that \object{AFGL 2298} is most luminous while at maximum radius, 
reversing the behaviour of the other LBVs.

Both terminal wind velocity and (unclumped) mass loss rates determined for 
\object{AFGL 2298} (Table~\ref{summary_afgl}) are comparable to those 
found for 
other highly luminous (candidate) LBVs such as the \object{Pistol Star} 
and \object{FMM~362} (Najarro et al. \cite{paco08}). The maximum mass-loss 
rate observed for \object{AFGL 2298} is significantly lower than that 
found for \object{$\eta$ Car} and, somewhat surprisingly, the cool B 
supergiant \object{HDE 316285}, which is of a similar temperature but is a 
factor of $\sim$5 less luminous (Hillier et al. \cite{hillier}, 
\cite{hillier98}). Both these stars support winds with very high 
performance numbers ($\eta$ (=c \.{M} v$_{\infty}$/L) of 18 and 4.5, 
respectively;  Hillier et al. \cite{hillier98}, \cite{hillier}); the lower 
value for \object{AFGL 2298} ($\eta\sim$unity) demonstrating that its 
current wind could likely be driven through radiation pressure. 
The 
results of our modelling -- in particular the comparatively modest mass 
loss 
rates --  indicate that the changes in physical properties of \object{AFGL 
2298} reflect real changes in the stellar temperature and radius.

\begin{figure}
\resizebox{\hsize}{!}{\includegraphics[angle=0]{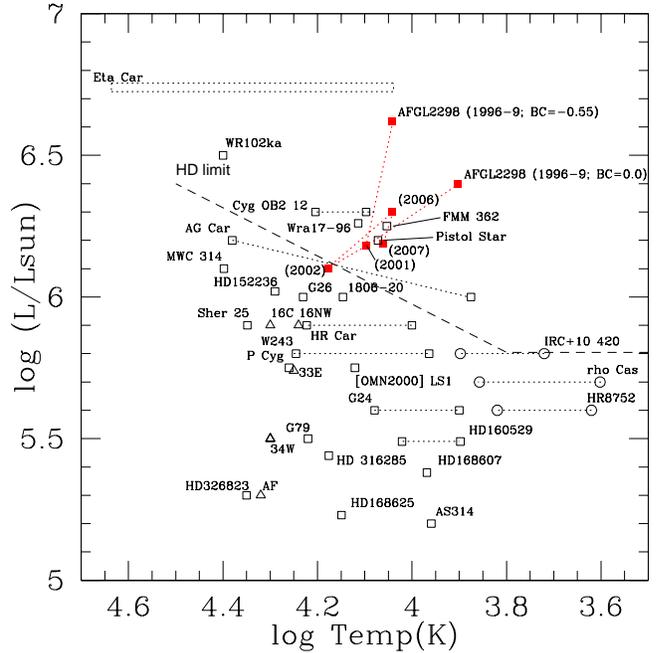}}
\caption{H-R diagram showing the positions of AFGL 2298 between 1996--2007 
in comparison to both 
confirmed  and candidate  Galactic LBVs (squares), narrow emission line sources from the Galactic Centre cluster
(triangles) and Galactic YHGs (circles). We present  two possible 
locations for AFGL 2298 between 1996--1998 based on 
temperatures of 8 kK and 11.5 kK (see Sect. 4 for justification). 
References for luminosity and temperature for 
individual systems are given in  Clark et al. (\cite{clark05}), with the exception of the Galactic Centre 
narrow emission line sources (Martins et al. \cite{martins}),
the \object{Pistol Star} and FMM362 (Najarro et al. \cite{paco08}),
\object{SGR 1806-20} (Bibby et al. \cite{bibby}),
\object{HD 326823} (Marcolino et al. \cite{marcolino}),
\object{AG Car} (Groh et al. \cite{groh}),
\object{HR Car} (Szeifert et al. in prep.)
\object{W243} (Ritchie et al. submitted),
\object{WR 102ka} (Barniske et al. \cite{barniske}),
\object{[OMN2000] LS1} (Clark et al. \cite{clark09}) and 
\object{G24.73+0.69} (Clark et al. in prep.).}\label{hrd}
\end{figure}

Finally, our spectroscopic modelling enables us to determine BCs 
for \object{AFGL 2298} as a function of stellar temperature and hence 
interpret the photometric lightcurve prior to 2001. In doing so, we find 
that assuming a BC$\sim$0 mag (corresponding to T$_{\ast}\sim8$ kK; 
appropriate for an LBV in the cool state)  during the 1996--1999 maximum 
results in log(L$_{\ast}$/L$_{\odot}$)$\sim$6.4, rising to 
log(L$_{\ast}$/L$_{\odot}$)$\sim$6.6 for an assumed T$_{\ast}\sim11$ kK, 
with the former estimate representing a robust lower limit to 
luminosity at  this time.{\em In comparison to the 2007 results, this 
demonstrates at a $>$5$\sigma$ level of significance that the L$_{\rm bol}$ 
of  \object{AFGL~2298} was not conserved between 1996--2007 }. We 
present an updated HR diagram in Figure~\ref{hrd}. At its highest 
luminosity,
\object{AFGL 2298} was one of the brightest stars in the Galaxy, being 
rivalled and surpassed only by \object{WR 102ka} (Barniske et al. 
\cite{barniske}) and $\eta$ Car (Hillier et al. \cite{hillier98}).

\begin{table*}
\begin{center}
\caption{Summary of the stellar properties of \object{AFGL 2998} and other 
LBVs observed to undergo giant eruptions.}\label{lbv_summary}
\begin{tabular}{lcccccccl}
\hline
\hline
Star & Epoch & $R_{\ast}$     & $T_{\rm eff}$ & log($L_{\ast}/L_{\odot}$) 
& 
H/He & $v_{\infty}$ 
& log $\dot M$ /$\sqrt f$ & Reference\\
     &      & ($R_{\odot}$)  & (kK)      &                           &      
& (kms$^{-1}$) & $M_{\odot}$ yr$^{-1}$ & \\
\hline
AFGL 2298 & 1996--9   &  -   & -   & $>$6.4 &  - &  - & - & a\\
              & 2001 Jun & 262 &  11.7 & 6.2 & {\em 5.0} & {\em 200} & 
-4.3 & a\\
              & 2002 Aug & 158 &  10.9 & 6.1 & {\em 5.0} & {\em 200} & 
-3.9 & a \\
              & 2006 Jun & 385 &  10.3 & 6.3 & {\em 5.0} & 125       & 
-4.3 & a \\
              & 2007 May & 312 &  10.9 & 6.2 & {\em 5.0} & 125       & 
-4.4 & a \\
\hline
NGC 2363-V1          &$<$1993    &  -   &  -    & $\leq${\em 5.9} &   - &    
-    &  -   &b \\             
                    &1997 Nov & 356 & 13.5   & 6.58& {\em 4.0} & 325       
& -3.3 & b,c\\
                    &1999 Jul & 201 & 18.5   & 6.63& {\em 4.0} & 290       
& -3.2 & b,c\\
                    &2000 Oct & 225 & 17.5   & 6.63& {\em 4.0} &  -        
& -3.3 & c, d\\
                    &2001 Dec & 223 & 18.5   & 6.72& {\em 4.0} &  -        
& -3.5 & c, d\\
                    &2003 May & 194 & 20.5   & 6.78& {\em 4.0} &  -        
& -3.8 & c, d\\
                    &2004 May & - &{\em 26.0}& {\em 6.80}&  -   &    -       
& {\em $<$ -3.4} &c, d \\
\hline 
HD 5980            &1978-80  &  -   &{\em 60.5}  & {\em 6.4}  &  - & - &  - 
& e\\
                    &1993 Nov &  -   &{\em 52.0}  & {\em 6.5}   & - & - & 
- & e\\
                    &1994 Sep & 280 & 23   & 7.05& 0.7       & 500       & 
-3.1 & b\\
                    &1994 Dec & 130 & 35   & 6.50& 2.5       & 600       & 
-3.0 & e\\
\hline      
$\eta$ Car & 1840's       &  -   & -     & {\em 7.5} &   -   
& {\em 650 - 3000}     & {\em $\ga$-0.3} & f, g\\
                      & 1890's       &  -   & -     &  -        &   -   & 
{\em 200}               &  {\em -1.8}    & h, i \\
                      & 1998.2      & 60-881  & 35.3-9.2 & 6.74 & 5.0 & 
500 & -3.0 & j\\ 
\hline
P Cygni       & 1600's       &  -   & -     & -  &   -   &     
{\em 137}           &{\em -1.7}    &k  \\
                        & 1989-92     &   76 & 18.7 &  5.8       &   3.5 & 
185 & -4.5 &l \\
\hline
\end{tabular}
\note{ Those 
parameters that are not directly determined from non-LTE model atmosphere analysis - for example wind parameters derived 
from present day nebular properties - are given in italics. Due to the extreme mass loss rate for 
\object{$\eta$ Car} a degeneracy is present for the stellar radius and temperature (Hillier et al. \cite{hillier}).
References used in the construction of this table are
$^a$ Clark et al. (\cite{clark03a}) and this study;
$^b$ Drissen et al. (\cite{drissen});
$^c$ Petit et al. (\cite{petit05});
$^d$ Petit et al. (\cite{petit});
$^e$ Koenigsberger et al. (\cite{koen})
$^f$ Humphreys et al (\cite{h99});
$^g$ Smith et al. (\cite{smith03});
$^h$ Smith  (\cite{smith05});
$^i$ Smith  (\cite{nature}); $^j$ Hillier et al. (\cite{hillier});
$^k$ Smith \& Hartigan (\cite{smith06b}); $^l$ Najarro (\cite{pacopcyg}). Volume filling factors, $f$, range from  
0.1 (\object{HD 5980} and \object{$\eta$ Car}),  through 0.3 (\object{NGC 
2363--V1}) to 0.5  (\object{P Cygni}). Time averaged mass 
loss  rates for \object{$\eta$ Car} and \object{P Cygni} derived from 
nebular masses are absolute (i.e. $f$=1). Finally, note that the values of $V_{\infty}$  presented  for 
NGC 2363-V1 in 1997 \& 1999 are from Drissen et al. (\cite{drissen}), while R$_*$, T$_{eff}$, log(L$_{*}$) and $\dot {M}$ 
are from the updated models presented by
Petit et al. (\cite{petit05}).}
\end{center}
\end{table*}

\section{AFGL 2298 in the context of LBV eruptions}
 
Our results strongly support the hypothesis that 
the  bolometric luminosity of \object{AFGL 2298} has varied by at 
least a 
factor of  $\sim$2 over the past 2 decades (1996--1999 compared to 
2002)\footnote{In the following discussion we shall assume that the 
properties 
of \object{AFGL 2298} in 2001--2002 are those of it in quiescence, despite 
this being significantly cooler than e.g. \object{AG Car} in such a state 
(24 kK;  Groh et al. \cite{groh}); if this is incorrect then the 
duration 
of the 
non-luminosity conserving behaviour and the magnitude of such changes 
will both be lower limits.}. Consequently, these variations differ from both 
the canonical excursions of LBVs to lower temperatures at near constant 
luminosity, {\em and} also the reduction in luminosities inferred for 
both 
\object{AG Car} and \object{S Dor} during such transitions to cooler 
states, for which the deficit in radiative energy is thought to be due to 
the requirement to support the expansion of the outer layers of the star.

Given that increases in L$_{\rm bol}$ for other LBVs have previously only 
been 
associated with \object{$\eta$ Car}-like giant eruptions, these 
observations are of considerable interest; more so given the twin 
suggestions that such behaviour is accompanied by significant mass loss - 
permitting the transition from Main Sequence to hydrogen depleted 
Wolf-Rayet and yielding the characteristic LBV ejection nebulae - and may 
presage the occurrence of a SN (Sect. 1). As such, does the `outburst' of 
\object{AFGL 2298} permit it to serve as a template system to calibrate 
other less well observationally constrained events, or are the known 
non-luminosity  conserving eruptions too heterogeneous to allow this?

\subsection{\object{HD~5980} and \object{NGC2363--V1}}

Prior to \object{AFGL 2298}, two LBVs have been the subject of
detailed spectroscopic analysis during the course of a giant eruption:
\object{HD 5980} in the SMC and NGC2363--V1  within NGC 2366, 
both low 
metallicity 
environments. Results of a synthesis of the long-term lightcurves, 
optical spectroscopy and, for a subset of the data, non-LTE model 
atmospheric analysis (Drissen et al.  \cite{drissen}, Koenigsberger et al. 
\cite{koen}) are presented in Table~\ref{lbv_summary}. A comparison 
between \object{AFGL 2298}, HD~5980 and NGC~2363--V1
reveals that the properties of the three outbursts differ 
significantly from one another in terms of the duration of the event - 
$\sim$6 months for \object{HD 5980} (e.g. Koenigsberger et al. 
\cite{koen}) 
to $>$7 yrs for \object{NGC 2363-V1} (Petit et al. \cite{petit}) and 
apparently \object{AFGL 2298}. Unless fortuitously located when 
\object{AFGL 2298} was inaccessible, the sampling frequency of the 
lightcurve would appear to exclude such a rapid outburst as that of 
\object{HD 5980}, although one could have been missed in \object{NGC 
2363--V1} (Petit et al. \cite{petit}).

Additionally, significant variance is found between peak luminosity and 
the evolution of stellar temperature during the outbursts 
(Table~\ref{lbv_summary}). In  particular, while the temperature of 
\object{AFGL 2298} varied by only  $\sim$4.5 kK between 2001--2007, the 
initial, pre-1997, phase of the  outburst of \object{NGC 2363--V1} must 
have been accompanied by a  significant reduction in temperature (for any 
reasonable progenitor;  Drissen et al. \cite{drissen}), a trend which has 
since reversed, with  temperature increasing by $\sim$13 kK over the 
following seven years, 
during which time L$_{\rm bol}$ has continued to increase.  Conversely, 
the increase in L$_{\rm bol}$ found for \object{HD 5980} appears to have 
been solely associated with a rapid and dramatic cooling 
($\Delta$T$\sim$30 kK), with the subsequent increase in 
temperature associated with a corresponding reduction in luminosity 
(noting that the long term secular changes in brightness prior to this 
event appear to have occurred at $\sim$constant L$_{\rm 
bol}$)\footnote{Our estimates of the stellar temperature of \object{HD 
5980}  prior to the 1994 June-October outburst were made using the 
photometric 
and spectroscopic summary present by Moffat et al. (\cite{moffat}) under 
the assumption that components B and C were of constant brightness 
(utilising values from Foellmi et al. \cite{foellmi}) and adopting 
temperatures and bolometric corrections appropriate for WN3h and WN6h 
stars in the Small Magellanic Cloud from Martins et al. 
(\cite{martins09}).}.

Finally, despite the lower metallicity of their host galaxies, the mass 
loss rates for both \object{HD 5980} and \object{NGC 2363-V1} are 
significantly higher than those observed for \object{AFGL 2298}, although 
for all three stars they appear to be orders of magnitude lower than those 
associated with \object{$\eta$ Car} and \object{P Cygni} in outburst 
(Table~\ref{lbv_summary}).

\subsection{The historic eruptions of \object{$\eta$ Car} and \object{P Cygni}}

Constraints on the outbursts of both stars may be inferred from their 
historical lightcurves and current properties of their circumstellar 
ejecta and are summarised in Table~\ref{lbv_summary}, supplemented with 
the results of non-LTE spectroscopic analyses of their current 
physical  properties. P  Cygni was observed to undergo two outbursts of 
$\sim$6yr 
and $\sim$3yr 
duration 
which commenced in 1600\,AD and 1655\,AD respectively. In both events it 
was 
inferred to brighten visually by 2-3 mag, with the colour information 
associated with the latter outburst consistent with a (non-unique) 
interpretation that the L$_{\rm bol}$ was also higher at this time (Lamers 
\& 
de Groot \cite{lamers92}). Smith \& Hartigan (\cite{smith06b}) report the 
discovery of a dynamically young 0.1 M$_{\odot}$ nebula associated with P 
Cygni which they attribute to the 1600\,AD event; if correct, this would 
imply a mass loss rate of $\sim$0.02 M$_{\odot}$yr$^{-1}$, $>$2 orders of 
magnitude larger than those found for \object{AFGL 2298}, \object{HD 5980} 
or \object{NGC 2363-V1}.

In a similar manner, Smith (\cite{smith05}) estimated a comparable time 
averaged mass loss rate for the 1890s outburst of \object{$\eta$ Car} 
(Table~\ref{lbv_summary}). The properties of the star during this 
outburst are difficult 
to define due to the uncertain circumstellar extinction, but Humphreys et 
al. (\cite{h99}) concluded that it is not necessary to postulate an 
increase in L$_{\rm bol}$ during this event. However, this was clearly not 
the case for the famous and well documented 1840s eruption 
(Table~\ref{lbv_summary} and 
references therein). While the $\sim$20 yr duration of the giant eruption of 
$\eta$ Car appears comparable to the ongoing eruption of \object{NGC 
2363--V1}
(Humphreys et al.  \cite{h99}), the stellar luminosity and mass loss rate 
were both extreme; the latter exceeding those of \object{AFGL 2298}, 
\object{NGC 2363--V1} and \object{HD 5980} by over 3 orders of magnitude.

 While the wind properties inferred for \object{AFGL 2298} are
relatively modest and in principle consistent with line driving, those
inferred for the 1840s and 1890s outbursts (and 
the 1600s outburst of \object{P Cygni}) are high enough that an 
alternative mechanism is required to power the outflows.  Moreover, based 
on the difference in properties between the two Homunculi and further 
supported by the recent discovery of a high velocity component of the 
1840s ejecta, Smith (\cite{smith05}, \cite{nature}) suggest that two 
different physical mechanisms may have initiated and driven the 1840s and 
1890s outbursts of \object{$\eta$ Car}.

\begin{table}
\begin{center}
\caption[]{Summary of the progenitor and peak 
bolometric luminosities of the extragalactic 
SNe imposters (top panel) and the dust enshrouded optical transients
 (lower panel).}\label{sn}
\begin{tabular}{lccl}
\hline
\hline
Transient &\multicolumn{2}{c}{ log(L$_{\ast}$/L$_{\odot}$)} & Reference  
\\
          &  Progenitor & Outburst \\        
\hline
SN 1961V & 6.4 & 8.7 &a, b\\
SN 1954J  & 5.7 & 6.5 &b\\
SN 1997bs & $<6.2$ & 7.4 &c \\
SN 2000ch  & 6.2  & 6.9 & d \\
\hline
NGC300-OT1  &4.5 & 6.6 & e\\
            & 4.7 & 7.1 & f\\
SN 2008S      &4.5 &  7.5 & g\\
 
\hline
\multicolumn{4}{l}{Luminosities assume a bolometric correction
of $\sim0$}\\
\multicolumn{4}{l}{as appropiate for an F supergiant. References
are}\\
\multicolumn{4}{l}{$^a$ Goodrich et al. (\cite{goodrich});
$^b$ Humphreys et al. (\cite{h99})}\\
\multicolumn{4}{l}{$^c$ Van Dyk et al. \cite{1997bs});
$^d$ Wagner et al. (\cite{wagner})}\\
\multicolumn{4}{l}{$^e$ Bond et al. (\cite{bond});
$^f$ Berger et al. (\cite{berger})}\\
\multicolumn{4}{l}{$^g$ Smith et al. (\cite{smith08b})}
\end{tabular}
\end{center}
\end{table}

\subsection{The extragalactic SNe imposters}

Finally, we consider the so-called SNe imposters or \object{$\eta$ Car} 
analogues. Humphreys et al. (\cite{h99}) and Van Dyk (\cite{VD}) review 
the nine examples identified prior to 2005; of these 4 have been little 
studied, while a fifth, \object{SN 2008kg} appears to be an LBV undergoing 
a normal LBV excursion and so is not discussed further (e.g. Maund et al. 
\cite{maund}). We summarise the limited information on the progenitor and 
peak bolometric luminosities of the remaining objects in Table~\ref{sn}. 
For  completeness, we also include two recently  discovered optical 
transients 
for which a connection with SNe imposters has been proposed (e.g. Bond et 
al. \cite{bond}, Smith et al.  \cite{smith08b}), although their dust 
enshrouded progenitors appear to be of lower mass than known LBVs.

As with \object{P Cygni} and \object{$\eta$ Car}, the lack of 
spectroscopic observations for the stars complicates determination of 
their quiescent and outburst properties; the values derived from the 
literature in Table~\ref{sn} typically assume an A/F spectral type in {\em 
outburst} (with the natures of the precursors summarised in Van Dyk et al. 
\cite{VD}, and references therein). However, we note the similarity of the 
outburst spectrum of \object{SN 1997bs} to those of \object{NGC2363--V1} 
circa 1997 (van Dyk et al. \cite{1997bs}, Petit et al. \cite{petit}), 
potentially implying an upwards revision for the outburst luminosity to 
log(L$_{\ast}$/L$_{\odot}$)$\sim$7.9.

If the dusty optical transients \object{SN2008S} and \object{NGC300-OT1} 
are super-Eddington outbursts (e.g. Smith et al. \cite{smith08}) rather 
than electron capture SNe (Thompson et al. \cite{thompson}), the mass 
range for such events will be extended downwards into the 
10--20M$_{\odot}$ 
regime (e.g. Berger et al. \cite{berger}). However, even without the 
inclusion of these objects, the progenitor and outburst luminosities of 
the SNe imposters emphasise the diversity of  non-luminosity 
conserving eruptions.

Conclusions over the timescale for such events are difficult to draw due to 
limited sampling of the lightcurves and possible observational biases 
introduced by surveys optimised to identify SNe. Nevertheless, the 
durations of the outbursts of \object{SN1961v}, \object{SN1954j} and 
\object{SN1997bs} ($<$1yr; Humphreys et al. \cite{h99} and refs. therein, 
van Dyk et al. \cite{1997bs})  all appear significantly shorter than those 
of \object{$\eta$ Car}, \object{NGC 2363--V1} and the behaviour of 
\object{AFGL 2298} at any stage since 1989. However the timescales of 
these outbursts, if not their magnitudes, are comparable to that of 
\object{HD 5980}. The variability of \object{SN2000ch} is characterised by 
it's extreme rapidity ($\sim$2mag. on timescales of a $\sim$week; Wagner et 
al. \cite{wagner}), with similar fluctuations also present during the 
1840s giant outburst of \object{$\eta$ Car} (Frew \cite{frew}); 
comparable 
behaviour appears to be absent for \object{AFGL 2298}, at least since 
2001.

\begin{table}
\begin{center}
\caption[]{Summary of the total mass and nebular expansion velocity of LBV (top panel) and YHG 
(bottom panel)  circumstellar ejecta.}\label{lbv_nebulae}
\begin{tabular}{lccl}
\hline
\hline
Star & v$_{neb}$ & M & Reference\\
     &  (km s$^{-1}$)& (M$_{\odot}$) \\
\hline
$\eta$ Car (Homunculus) & 650 &$>$12 &a \\
\hspace{0.7cm} (Little Homunculus) & 200 & 0.1 & b\\
P Cygni & 136 & 0.1  & c\\
Pistol Star & 60-95 & 11 & d, e  \\
AFGL 2298 & 70 & 10 & f  \\
AG Car & 70 & 8.9 & g, h\\
G79.29+0.46 & 44 & 15 & i, j, k\\
Wra 751 & 26 & 1.7 & g, l \\
HD 168625 & 19 & 2.1 & m, n\\
\hline
IRC +10 420  &  27 & 0.24 & o   \\  
             &  35 & 0.46 & o   \\
HD 179821   &  35 & 4.0 & o   \\
\hline
\multicolumn{4}{l}{$^a$Smith et al. (\cite{smith03});
 $^b$Smith (\cite{smith05})}\\
\multicolumn{4}{l}{$^c$Smith \& Hartigan (\cite{smith06b});
$^d$Figer et al. (\cite{figer99})}\\
\multicolumn{4}{l}{ $^e$Figer et al. (\cite{figer98});
 $^f$Ueta et al. (\cite{ueta})}\\
\multicolumn{4}{l}{ $^g$Voors et al. (\cite{voors});
 $^h$Smith et al. (\cite{smith91})}\\
\multicolumn{4}{l}{ $^i$Higgs et al. (\cite{higgs});
$^j$Waters et al. (\cite{waters})}\\
\multicolumn{4}{l}{ $^k$Voors et al. (\cite{voorsb});
 $^l$Hutsemekers \& van Drom (\cite{hut91})}\\
\multicolumn{4}{l}{ $^m$Pasquali et al. (\cite{anna});
 $^n$O'Hara et al. (\cite{ohara})}\\
\multicolumn{4}{l}{ $^o$Castro-Carrizo et al. (\cite{castro})}
\end{tabular}
\end{center}
\end{table}

\section{Outbursts and nebular formation}

Given the apparently diverse nature of giant eruptions described in 
Sect.~5, it is of interest to determine whether the 
physical properties of LBV ejection nebulae are similarly heterogeneous.  
In Table~\ref{lbv_nebulae} we summarise the properties of those nebulae 
surrounding {\em  Galactic} LBVs for which both current expansion velocity 
and mass have 
been determined (with an expanded summary presented in Clark et al. 
\cite{clark03b}). In compiling this we note that the nebulae associated 
with \object{P Cygni} and \object{$\eta$ Car} are relatively youthful 
($<400$yrs) compared to the dynamical ages of the remaining LBVs 
($>$10$^3$yr; references given in caption, Table~\ref{lbv_nebulae}); this 
will be returned to shortly.

One must also be careful to distinguish those stars which might have 
ejected their nebulae during a RSG phase, an issue highlighted by Voors 
et al. (\cite{voors}), who found that the dust composition of RSG ejecta 
is comparable to that of LBV nebulae, suggesting similar conditions during 
their formation. Assuming evolution at constant $L_{\rm bol}$ once stars 
enter 
the supergiant phase, both observational constraints and theoretical 
predictions suggest that stars with log(L$_{\ast}$/L$_{\odot}$)$\leq$5.8 
will pass through a Red Supergiant(RSG) /Yellow Hypergiant (YHG) phase 
prior to evolving to higher temperatures (Humphreys \& Davidson \cite{HD}, 
Meynet \& Maeder \cite{meynet}). Thus, from the stars listed in 
Table~\ref{lbv_nebulae},  \object{G79.29+0.46}, \object{Wra 751} and 
\object{HD 168625} could all have passed through such a phase.

Given such an hypothesis, it is notable that the mass and velocities of 
the ejecta associated with the YHGs \object{IRC +10 420} and \object{HD 
179821}, which are thought to be post-RSG objects, are comparable to those 
of \object{Wra 751} and \object{HD 168625} (although the mass of 
\object{G79.29+0.46} is significantly higher; Table~\ref{lbv_nebulae}). 
Assuming that the  YHG nebulae continue to expand at their current 
velocities, in 
$\sim$5000~yr they would be the same size (and age) as the \object{Wra 
751} nebula is today.  Additionally, \object{IRC +10 420} currently 
appears to be evolving to higher temperatures and an early Blue 
Supergiant/WNL phase (Oudmaijer \cite{rene}). At such a point the combined 
nebulae+stellar system would effectively be indistinguishable from 
\object{Wra 751}. Indeed observations of early Blue Supergiant/WNL stars 
in Wd1 - which have passed through a YHG phase - show them to be 
significantly variable as well (Clark et al. in prep.), thus replicating 
all observational properties of an LBV.
Therefore, it is at least possible that the ejecta around (a subset of) 
the low luminosity LBVs was formed in a previous cool hypergiant 
phase\footnote{See Smith (\cite{smith07}) for a counterargument 
for 
\object{HD 168625}.}. In this respect the 2001 outburst of \object{$\rho$ 
Cas} which resulted in a brief phase of extreme mass loss with 
\.{M}$\sim$10$^{-3}$M$_{\odot}$yr$^{-1}$ - directly comparable to the time 
averaged mass loss rates inferred for \object{IRC +10 420} and \object{HD 
179821} (Castro-Carrizo et al. \cite{castro}) - may form a workable 
paradigm.

Of the stars that are too luminous to have evolved through a RSG phase, 
there is a notable similarity in the properties of the nebulae of 
\object{AFGL 2298}, \object{AG Car} and the \object{Pistol Star} (Table~ 
\ref{lbv_nebulae}), while the stars themselves are all of comparable 
luminosity; highly suggestive of a uniform formation history.
In comparison to \object{P Cygni} and \object{$\eta$ Car} (during the 
production of the Little Homunculus), it is  unclear whether the 
differences in the nebular expansion velocities observed for \object{AFGL 2298}, 
\object{AG Car} and the \object{Pistol Star}
and, for the first two stars, the {\em time 
averaged} mass loss rates (Sect. 3.2, Voors et al. \cite{voors}) are 
solely due to observing the nebulae at different evolutionary stages. 
However, it does appear difficult to reconcile the extreme physical 
properties present during the formation of the Homunculus - in particular 
the velocity and mass loss rate - with those of \object{AFGL 2298}, 
\object{AG Car} and the \object{Pistol Star}.  

For example, Langer (2008, 
priv. comm.) suggested that after several thousand years the rapid 
expansion of the polar lobes of the Homunculus would reduce their surface 
brightness below our current detection threshold, leaving the slower 
moving equatorial material as the sole detectable result of the 1840s 
outburst, which by that time would resemble the nebulae around 
\object{AFGL 2298} et al. However, Smith et al. (\cite{smith03}) found 
that only 10-20\% ($\leq$2.5M$_{\odot}$) of the total mass of the 
Homunculus is located in the equatorial region, whereas the masses of the 
nebulae in question are $\sim$8.9-11M$_{\odot}$. Thus, at the very least 
it would appear that the latitudinal distribution of nebular mass would 
have to differ between these objects; applying the `equatorial' to `polar' 
mass ratio found for the Homunculus to \object{AFGL 2298}, \object{AG Car} 
and the \object{Pistol Star} would imply total nebular masses in excess of 
$\sim$50M$_{\odot}$.

\section{Concluding Remarks}

The continued spectroscopic and photometric monitoring of \object{AFGL 
2298} clearly indicates that it has been highly variable over the last 
$\sim$20~yr. The magnitude of variability ($>$1.6 mag in the near-IR) is 
typical of that observed for L$_{\rm bol}$ conserving outbursts in LBVs. 
However, the results of non-LTE model atmosphere analysis suggest that 
unlike normal S Dor-like LBV excursions, which occur at constant 
luminosity, the L$_{\rm bol}$ of \object{AFGL 2298} varied during this 
time, being a factor of $\geq$2 greater during 1996--1999 in comparison to the 
photometric minimum of 2001--2002 (indeed, during its 1996--1999 peak it 
was one of the most luminous stars known in the Galaxy). The changes in luminosity 
appear to be driven by expansion and contraction of the photosphere at 
$\sim$constant temperature. As such they differ from those observed for 
\object{S Dor} and \object{AG Car}, where an expansion of the star is 
accompanied by a significant cooling, resulting in an overall reduction in 
L$_{\rm bol}$. Throughout these changes both the wind velocity and mass 
loss 
rate of \object{AFGL 2298} were moderate, directly comparable with other 
highly luminous (candidate) LBVs such as the \object{Pistol Star} and 
\object{FMM~362} (Najarro et al. \cite{paco08}) and entirely consistent 
with expectations for a line driven wind.

By comparison to the 2001--2002 minimum, the excess energy radiated during 
the 
1996--1999 photometric maximum requires the star to generate an additional 
$\sim$10$^{47}$ergs. We note that with the exception of the Homunculus, this is
 $\sim$comparable to the current kinetic energies of LBV ejecta (determined from 
the nebular masses and expansion velocities given in 
Table~\ref{lbv_nebulae}). This value does not include the extra requirement of supporting the 
extended outer layers of the  star against gravity during this period. Without 
a measurement of the mass of material involved, it is difficult to infer 
the energy budget for this, but simply assuming a similar mass to that  
suggested for \object{S Dor} and \object{AG Car} (0.2--0.6M$_{\odot}$; 
Lamers \cite{lamers95}, Groh et al. \cite{groh}) would require the 
star 
to find 
a further $\sim$10$^{47}$ergs over this period.

Due to the renewed interest in the role that non L$_{\rm bol}$ conserving 
'Giant' LBV eruptions play in the evolution of massive stars and also 
their possible role in signalling incipient SNe, we compared the 
properties of \object{AFGL 2298} to those of other LBV eruptions. From the 
limited data available we found that the properties of such eruptions 
appear to be highly heterogeneous, to the extent that it was not possible 
to identify a characteristic template for such events. Timescale varied 
from $<$0.5 - $\sim$20~yr, with peak luminosities and mass loss rates 
spanning $>$2 and $>$3 orders of magnitude respectively.  Placed in this 
context, we found that the behaviour of \object{AFGL 2298} defined a lower 
bound to the properties of LBV eruptions in terms of increase in 
L$_{\rm bol}$ 
and mass loss rate.

Likewise, significant variation was found between the {\em current} 
properties of the ejection nebulae associated with LBVs. While this likely 
partly reflects observational selection effects introduced by observing 
nebulae at different stages in their life cycle, we suspect that real 
differences in the physics of the ejection event may also be involved. For 
example, a subset of the lower luminosity 
(log(L$_{\ast}$/L$_{\odot}$)$<$5.8) stars could eject their nebulae in 
a prior cool hypergiant, rather than the current LBV,  phase (e.g. Voors et al. \cite{voors}).
We note that a comparison of the disparate properties of the Little and Big Homunculi 
of \object{$\eta$ Car} also led Smith (\cite{smith05}) to suggest that the 
underlying physical cause of both events differed from one another. 
Indeed, the mass loss rate of \object{$\eta$ Car} during the production of 
the Homunculus is unrivalled by that of any other LBV, with the exception 
of the mass loss inferred for the progenitors of \object{SN2006gy} 
(Agnoletto et al. \cite{ag}) and \object{SN2006tf} (Smith et al. 
\cite{smith08}).

As such, given the diverse nature of the outbursts, nebulae and, where 
quantifiable, progenitors, one might consider the possibility that more 
than one physical process may result in non L$_{\rm bol}$ conserving 
eruptions 
(e.g. Smith \cite{smith05}), depending on the properties of the underlying 
(binary?) star. Indeed, a number of different mechanisms have already been 
proposed to explain Giant eruptions, including hydrodynamical 
instabilities (Smith \& Owocki \cite{smith06}), the pulsational pair 
instability (Woosley et al. \cite{woosley}), tidal interaction in a binary 
(Koenigsberger \cite{k08}) and binary mergers (Morris \& Podsialowski 
\cite{morris}). If this is the case, then the term Luminous Blue Variable 
would be more applicable as a phenomenological description of the 
observational properties of a star, rather than as a description of a 
unique evolutionary phase.

\begin{acknowledgements}

JSC acknowledges support from an RCUK fellowship. AZT-24 observations are 
made within an agreement between Pulkovo, Rome and Teramo observatories.

\end{acknowledgements}

{}
\end{document}